# Modeling and Propagation of Noisy Waveforms in Static Timing Analysis

Shahin Nazarian, Massoud Pedram
*Dept. of EE-Systems, University of Southern California*
*Los Angeles, CA 90089*

Emre Tuncer, Tao Lin, Amir H. Ajami
*Magma Design Automation*
*Santa Clara, CA 95054*

## Abstract

*A technique based on the sensitivity of the output to input waveform is presented for accurate propagation of delay information through a gate for the purpose of static timing analysis (STA) in the presence of noise. Conventional STA tools represent a waveform by its arrival time and slope. However, this is not an accurate way of modeling the waveform for the purpose of noise analysis. The key contribution of our work is the development of a method that allows efficient propagation of equivalent waveforms throughout the circuit. Experimental results demonstrate higher accuracy of the proposed sensitivity-based gate delay propagation technique, SGDP, compared to the best of existing approaches. SGDP is compatible with the current level of gate characterization in conventional ASIC cell libraries, and as a result, it can be easily incorporated into commercial STA tools to improve their accuracy.*

## 1 Introduction

STA tools require accurate delay models for gates and interconnects. The function of a *gate delay model* is to take a (noisy) input waveform at the far-end of the interconnect (e.g., *in_u* in Figure 1) and produce the waveform for the gate output (*out_u.*) This process is known as the *gate delay propagation*. Most STA tools model the noisy input waveform at *in_u* with an equivalent linear waveform, $\Gamma_{eff}$, with a single reference point (input arrival time) and a constant slope (an effective input slew.) However, different waveforms with identical arrival time and slew when applied at *in_u* can result in very different propagation delays through *4INV$_x$*. Generally speaking, as the crosstalk noise becomes more significant in current technologies, using only a reference point and a constant slope to convey the timing information for a signal transition adversely impacts the accuracy of STA tools. More precisely, the actual shape of the input waveform should be considered to ensure accuracy.

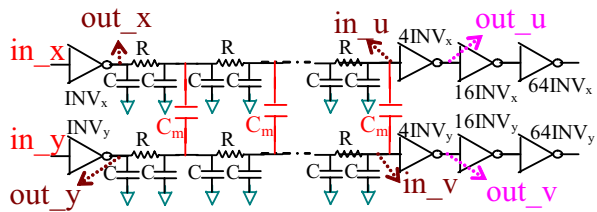

**Figure 1. Our experimental setup. R=8.5Ω, C=4.8fF.**

## 2 Conventional Gate Delay Propagation Techniques

### 2.1 Point-Based

A technique, denoted by **P1**, sets the input slew of $\Gamma_{eff}$ to the time from the $0.1V_{dd}$ to $0.9V_{dd}$ crossing points of the noiseless waveform as though the waveform had not been affected by the noise. Another technique, called **P2**, uses the time from the earliest $0.1V_{dd}$ crossing point to the latest $0.9V_{dd}$ crossing point of the noisy waveform as the effective slew of $\Gamma_{eff}$. Both techniques set the $0.5V_{dd}$ point of the $\Gamma_{eff}$ to the latest $0.5V_{dd}$ crossing point. P1 and P2 may be too optimistic in some cases because of the way that they calculate the slew of $\Gamma_{eff}$. They can also be too pessimistic in other cases because of the way they calculate the arrival time.

### 2.2 Least Squared Error-Based

A technique, denoted by **LSF3**, finds $\Gamma_{eff}$ such that the sum of the squares of the sampled differences between $\Gamma_{eff}$ and the input noisy waveform is minimized. LSF3 can show pessimistic or optimistic behavior because it is simply a mathematical approach to match a waveform without any consideration of the logic gate behavior.

### 2.3 Energy-Based

Inspired by the Elmore delay idea [2], another technique, denoted as **E4**, passes $\Gamma_{eff}$ through the latest $0.5V_{dd}$ crossing point of the noisy voltage waveform. The slope is then selected such that the area which is encapsulated by that line, and straight lines $v_1(t) = 0.5V_{dd}$ and $v_2(t) = V_{dd}$ is equal to the area that is enclosed by the noisy input and lines $v_1$ and $v_2$. In general, the more times the noisy waveform passes through the $0.5V_{dd}$ level, the higher is the probability for this approach to produce pessimistic estimates.

### 2.4 Weighted Least Squared Error-Based

Recently, a technique, which we will denote as **WLS5**, has been suggested in [1] that multiplies a weight factor to each squared term in the minimization equation of LSF3. This factor is defined as the derivative of the output to input signal for the noiseless input.

**WLS5-Step 1: Finding the derivative of the output to the noiseless input.** For each logic cell, the derivative of the output to the noiseless input waveform, $\rho^{noiseless}$, is calculated as:

$$\rho^{noiseless}(t) = \partial v_{out}^{noiseless}(t) \Big/ \partial v_{in}^{noiseless}(t) = \frac{\partial v_{out}^{noiseless}(t)/dt}{\partial v_{in}^{noiseless}(t)/dt} \quad (1)$$

where $v_{in}^{noiseless}(t)$ and $v_{out}^{noiseless}(t)$ are the noiseless input and its resulting output voltage values at time $t$, respectively. Note that $\rho^{noiseless}(t)$ is equal to the ratio of output slew to noiseless input slew (Figure 2.a.) This weight factor is non-zero only for points in a critical region, called the noiseless critical region, and is considered to be zero outside that region. The noiseless critical region is defined between $t_{first}^{noiseless}$ and $t_{last}^{noiseless}$, which are in turn set to time instances at which the noiseless input crosses the $0.1V_{dd}$ and $0.9V_{dd}$ levels, respectively.

**WLS5-Step 2: Finding $\Gamma_{eff}$.** WLS5 finds $\Gamma_{eff}$ with coefficients $a$ and $b$ such that Equation 2 is minimized ($P$ denotes the number of sampling points in this paper):

$$\sum_{k=0}^{P-1} \{\rho^{noiseless}(t_k)(v_{in}^{noisy}(t_k) - (a \times t_k + b))^2\} \quad (2)$$

The noiseless critical region in WLS5 acts as a filter. If the noise distortion occurs outside the noiseless critical region, then it will be ignored. Our experiments confirm that limiting the noise consideration to this range only, makes WLS5 inaccurate. Moreover, the higher the number of aggressors is, the higher is the probability that WLS5 underestimates the arrival time and/or slew at the output of the gate by a large amount.

Another weak point of WLS5 is that it is meaningful as long as the noiseless input and output waveform overlap each other; otherwise, the derivative of output to input is undefined. Therefore, WLS5 cannot be applied to gates with large intrinsic delay such as multi-stage gates, and/or those with large fanout loadings, where the input and output transitions may not overlap.

## 3 SGDP: Sensitivity-Based Gate Delay Propagation

This section describes the main steps of the proposed SGDP.

### 3.1 Calculation Steps

**SGDP-Step 1: Finding the derivative of the output to the noiseless input.** This step is the same as that in WLS5.

**SGDP-Step 2: Estimation of the derivative of the output to the noisy input.** This step produces an approximation of $\rho^{eff}$, which denotes the derivative of the output with respect to the noisy input voltage waveform. $t_{first}^{noisy}$ and $t_{last}^{noisy}$ are defined to delimit the critical region of the noisy waveform, i.e., they are set to time instances at which the noisy input ($v_{in}^{noisy}$) crosses the $0.1V_{dd}$ for the first time and the $0.9V_{dd}$ for the last time, respectively. $\rho^{eff}$ is calculated from $\rho^{noiseless}$ as follows: *2.a) For every $t_i \in [t_{first}^{noisy}, t_{first}^{noisy}]$, find $t_j \in [t_{first}^{noiseless}, t_{last}^{noiseless}]$ such that: $v_{in}^{noisy}(t_i) = v_{in}^{noisy}(t_j)$, 2.b) Next set $\rho^{eff}(t_i) = \rho^{noiseless}(t_j)$.* More precisely, at each time level in the range $[t_{first}^{noisy}, t_{last}^{noisy}]$, for each voltage level, $\rho^{eff}$ is set to the corresponding derivative from the noiseless waveform at the same input voltage level. In this way, SGDP can consider the noise distortion in the noisy critical region. This overcomes the first shortcoming of WLS5, which would distortion if it occurred outside the noiseless critical region.

**SGDP-Step 3: Finding $\Gamma_{eff}$.** SGDP next finds $\Gamma_{eff}$ with coefficients $a$ and $b$ such that $\Delta_{out}$, the sum of the squares of the sampled differences between the resulting output ($v_{out}^{eff}$) and the actual output ($v_{out}^{noisy}$) is minimized. Equation 3 is an approximation of $\Delta_{out}$ using the first two terms of Taylor series:

$$\sum_{k=0}^{P-1}\{\rho^{eff}(t_k) \times (v_{in}^{noisy}(t_k) - (a \times t_k + b)) + \frac{1}{2} \times \frac{\partial \rho^{eff}(t_k)}{\partial v_{in}^{noisy}(t_k)} \times (v_{in}^{noisy}(t_k) - (a \times t_k + b))^2\} \quad (3)$$

Figure 2.b depicts $\rho^{eff}$, $\Gamma_{eff}$, and $v_{out}^{eff}$ as calculated by SGDP.

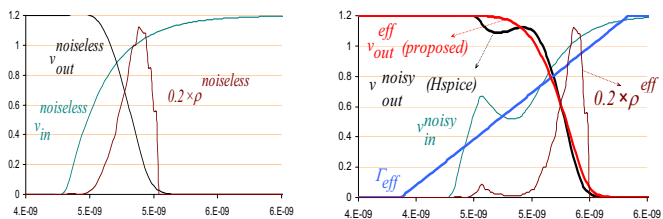

**Figure 2. Waveforms of a) $\rho^{noiseless}$, and b) $\rho^{eff}$, $\Gamma_{eff}$, and $v_{out}^{eff}$.**

To address the weakness of WLS5 for gates with non-overlapping input and output voltage transitions, SGDP adds additional pre- and post-processing steps as follows:

**SGDP-Additional step for non-overlapping input and output waveforms only:** SGDP shifts the output back in time by an amount $\delta$ such that $0.5V_{dd}$ for both the input and output voltage waveforms coincide. It then performs SGDP-Steps 1, 2, and 3. Finally, it shifts the equivalent input line forward in time by $\delta$.

## 4 Experimental Results

### 4.1 Accuracy Comparison

Table 1 shows the gate delay errors for all of the techniques discussed in this paper, including SGDP, compared to Hspice. The gate delay was calculated as the difference between the $0.5V_{dd}$ crossing points of the input and output waveforms.

Configuration I is the one depicted in Figure 1 with total coupling value of 100fF. We used standard inverter cells of an industrial TSMC $0.13\mu$ cell library in our experiments. Both aggressor and victim line inputs, *in_x* and *in_y*, were 1000µm long and were given a slew of 150ps. Configuration II includes two aggressors *x1* and *x2* each with 100fF total coupling capacitance. These aggressors and the victim line, *y*, were each 500µm long and modeled similarly to the interconnects in Figure 1. 200 noise injection timing cases in a range of 1ns were analyzed for each configuration. Results are reported in Table 1. SGDP shows higher accuracy compared to all existing techniques. For instance, for configuration II, the average (maximum) delay error reduction is 2.6ps (4.8ps) compared to WLS5, which is the most accurate technique among the conventional ones. Therefore SGDP reduces the average (maximum) delay error by %15 (%10) over WLS5.

**Table 1: Accuracy comparison among all techniques.**

| Method | Delay Error (*ps*) | | | |
| --- | --- | --- | --- | --- |
| | Configuration I | | Configuration II | |
| | Max | Avg | Max | Avg |
| P1 | 81.3 | 29.3 | 134.2 | 48.5 |
| P2 | 82.7 | 24.5 | 144.5 | 51.3 |
| LSF3 | 75.1 | 30.9 | 110.8 | 45.4 |
| E4 | 82.3 | 14.5 | 145.3 | 33.4 |
| WLS5 | 42.4 | 10.3 | 49.3 | 17.4 |
| **SGDP** | 38.3 | 9.2 | 44.5 | 14.8 |

### 4.2 Run-Time Comparison

Although the worst case computational complexity of all techniques including the SGDP can be proved to be of linear order with respect to $P$, in practice, we observed different run times. On average P1, P2, and LSF3, and E4 take about 40µs and WLS5 takes about 60µs to accomplish delay propagation through a typical logic gate on a *Sun Blade 1000* machine. In contrast, both WLS5 and SGDP (with $P = 35$) take about 65µs. The SGDP run-time can be reduced by using smaller $P$ values. However small $P$ tends to result in lower timing analysis accuracy.

## 5 Conclusions

This paper presented a technique to efficiently propagate gate delay information for noisy waveform for the purpose of STA. Without any additional library characterization, it utilizes the sensitivity of the output to the noisy input waveform to model the impact of the shape of the waveform. Experiments demonstrate that this technique is more accurate than any existing technique e.